\documentclass[twocolumn,           
               showpacs,            
               nopreprintnumbers,     
               aps,                 
               prd,          	    
               letterpaper,             
              groupeaddress,      
               nofootinbib,         
               tightenlines,        
               floats,floatfix,      
               showkeys
               ]{revtex4-1}
               
\usepackage[toc,page]{appendix}
\usepackage{graphicx}
\usepackage{dcolumn}
\usepackage{bm}
\usepackage{amsmath}
\usepackage{amsfonts,amssymb}
\usepackage{soul}
\usepackage{color}
\definecolor{v}{rgb}{0.6, 0.2, 0.8} 
\definecolor{MAGA}{rgb}{0.1, 0.43, 0.75}
\definecolor{jm}{rgb}{0.13, 0.48, 0.64}

\usepackage{xcolor}
\usepackage{enumerate}
\usepackage{float}
\usepackage{subfigure}
\usepackage{ulem}

\usepackage{tabularx}

\begin{document}

\title{MeV Dark Energy Emission from a De Sitter Universe}

\author{Yasm\'in B. Alc\'antara-P\'erez$^{1}$}
\email{yasbe21@yahoo.com.mx}
\author{Miguel A. Garc\'ia-Aspeitia$^{2}$}
\email{angel.garcia@ibero.mx}
\author{Humberto Mart\'inez-Huerta$^3$}
\email{humberto.martinezhuerta@udem.edu}
\author{A.  Hern\'andez-Almada$^4$}
\email{ahalmada@uaq.mx}

\affiliation{$^1$Departamento de F\'isica, DCI, Campus Le\'on, Universidad de Guanajuato, Le\'on 37150, Guanajuato, Mexico}
\affiliation{$^2$Depto. de Física y Matemáticas, Universidad Iberoamericana Ciudad de México, Prolongación Paseo \\ de la Reforma 880, México D. F. 01219, México.}
\affiliation{$^3$Departamento de Física y Matemáticas, Universidad de Monterrey,
Av. Morones Prieto 4500, San Pedro Garza Garc\'ia, Nuevo León, 66238, Mexico}
\affiliation{$^4$Facultad de Ingenier\'ia, Universidad Aut\'onoma de 
Quer\'etaro, Centro Universitario Cerro de las Campanas, 76010, Santiago de
Quer\'etaro, M\'exico,}

\begin{abstract}
The evolution of a de Sitter Universe is the basis for both the accelerated Universe and the late-stationary Universe. So, how do we differentiate between both universes? In this paper, we state that it is not possible to design an experiment using luminous or angular distances to distinguish between the two cases because they are the same during the de Sitter phase. However, this equivalence allows us prediction of the signal of {\it a constant dark energy emission} with a signal peak around $29.5$ MeV, in which, according to our astrophysical test of survival probability, the radiation must be {\it non-standard photons}. Remarkably, experiments by EGRET and COMPTEL have observed an excess of gamma photons in this predicted region, coming from a possible decay process of  dark energy emission, which may constitute the smoking gun of a late-stationary Universe with the continuous creation of non-standard radiation, an alternative approach to understanding the current stages of the Universe's evolution.
\end{abstract}
\pacs{}
\maketitle

\section{Introduction}
The late acceleration of the Universe---which was first observed by the teams dedicated to study the supernovae Type Ia (SNIa)~\cite{Riess:1998,Perlmutter:1999} and was later confirmed by the WMAP and Planck satellites through cosmic microwave background radiation (CMB)~\cite{Planck:2018}---is today an undisputed fact. According to the general theory of relativity (GR), and assuming a homogeneous and isotropic line element and a perfect fluid energy-momentum tensor, the scale factor for an accelerated Universe ($z\lesssim0.6$) is expressed as $a(t)=a_0\exp[\Lambda(t-t_0)]$, where $\Lambda$ is the cosmological constant at $z=0$, representing a de Sitter evolution. Additionally, the equation of state (EoS) of the fluid responsible for the acceleration must fulfill inequality $\omega<-1/3$. 
According to these demands, the cosmological constant (CC) is one of the explanations, and through this approach, it is possible to construct the standard paradigm for cosmology, also known as the $\Lambda$ cold dark matter ($\Lambda$CDM) model, which is in agreement with modern observations. However, we should not lose sight of the profound unsolved problems afflicting the CC~\cite{Zeldovich,Weinberg}, which can be attributed to possible modifications to the GR or to a misinterpretation of the CC from the quantum field theory point of view, in which the ultraviolet divergence of the energy density arises. 

From the experimental approach, we know that we are in an accelerated phase because the observed luminous distances from SNIa (and confirmed by CMB~\cite{Planck:2018}) coincide with the theoretical expression expected for an accelerated Universe according to the background model. The luminous distance expression is  $d_L(z)\approx H_0^{-1}\Omega_{\Lambda}^{-1/2}(1+z)z$, where $\Omega_{\Lambda}$ is the density parameter of the CC, and it is valid only when $\Omega_{\Lambda}$ dominates over the other components (matter and radiation) \footnote{This assumption is valid because $\Omega_{\Lambda}=0.68$ according to~\cite{Planck:2018}, and eventually $\Omega_{\Lambda}=1$, while the other components tend to dilute.}. 
A similar situation happens, for example, with other observations, like Baryon Acoustic Oscillations (BAO), when we explore the final stages of the Universe's evolution, where $\Omega_{\Lambda}$ dominates. In this case, the angular distance can be reduced to \mbox{$d_A(z)\approx  zH_0^{-1}\Omega^{-1/2}_{\Lambda}(z+1)^{-1}$}. This confirms that the observations and theory fit only if the Universe is considered to be in an accelerated stage nowadays.
Both these measurements (and others) are important evidence that the Universe is transiting to a de Sitter phase in its last stages, and two of its main physical observables are the luminous and angular distances.

On the other hand, the steady-state model (SSM) demands a {\it perfect cosmological principle} (PCP)~\cite{BondiandGold}, generalizing the isotropy not only in all directions at cosmological scales but also at all times. Indeed, the SSM demands $\dot{a}/a=H_0$, where $H_0$ is the Hubble constant, implying $a=a_0\exp[H_0(t-t_0)]$ with a deceleration parameter $q\simeq-1$ and a luminous and angular distance that coincides precisely with the one obtained from the de Sitter evolution in the $\Lambda$CDM model in its last stages.
We note that the SSM was initially discarded because of its inability to predict CMB radiation and its black-body spectrum~\cite{weinberg1972gravitation}. 
Additionally, when the SSM was proposed, there was no observational evidence regarding the transition of our Universe into an accelerated stage; thus, the best at that epoch was an universe dominated by matter at $z\sim0$, which is no longer the case today. Nowadays, SSM has evolved into the so-called {\it matter/radiation creation model }\cite{Trevisani:2023wpw,Cardenas:2020exv,Lima:2015xpa,VargasdosSantos:2014ddg,Ramos:2014dba}, in which the de Sitter evolution is not caused by a cosmological constant; instead, a continuous creation of matter/radiation is produced through a diffusion term in the continuity equations driving the de Sitter evolution observed.

Evidence shows that our Universe is an evolving system rising from a Big Bang 13.7~billion years ago, with different transitions, but in particular with a transition at $z\lesssim0.6$ that coincides with the mathematical description of both an accelerated and a {\it late steady-state universe} (LSS). Additionally, based on the $\Lambda$CDM model, the Universe eventually tend to $q\to-1$ when $z\to-1$, i.e., in the far future.
The word {\it late} is used because the steady state is not valid for all epochs in the Universe evolution; prior  to this epoch (accelerated/stationary), the $\Lambda$CDM model is still the cornerstone. This affirmation arises from the mathematical equivalence between models. Thus, how can we be certain of which model for the Universe is the ideal interpretation for $z\leqslant0.6$? Are both conditions equivalent, and is there no way to differentiate between an accelerated and a steady-state Universe for $z\leqslant0.6$? If there is an equivalence, maybe the interpretation under one condition is easier than under the other (accelerated $\leftrightarrow$ LSS), as it happens with the equivalence principle or the equivalence between anti-de Sitter and the conformal field theory (AdS/CFT).

Thus, this paper is dedicated to tackling these questions and following up on all of their consequences. The outline of the paper is as follows: Section~\ref{LSS} is dedicated to discussing the LSS model; it presents the predictions related to the continuous emission of radiation at MeV energy. Section~\ref{Sec:AstroTests} tackles the consequences of MeV emission at astrophysical scales and concludes that this radiation must belong to the dark sector. In Section~\ref{Sec:FieldTheory}, the underlying classical field theory and the consequences in cosmology are revisited.  Finally, in Section~\ref{DC}, we present our discussions and conclusions.

\section{The LSS Model}\label{LSS}

LSS requires the continuous creation of matter and radiation in order to have a steady-state condition, implying the violation of the conservation of the energy-momentum tensor  $\nabla^{\mu}T_{\mu\nu}\neq0$, which is incompatible with GR and, thus, we need to have matter creation at a rate of $\sim3H$ per the existing accumulation of matter in the Universe, i.e., we need to maintain $\dot{\rho}=\dot{p}=0$ for the dominant components. On the other hand, we note that the CC shows similar complications: in this case, we also need to fulfill  condition $\dot{\rho}_{\Lambda}=\dot{p}_{\Lambda}=0$, pushing GR to its limits due to the continuous creation of energy/matter to keep CC constant. 
\footnote{This is a consequence of introducing quantum--vacuum fluctuations to account for this continuous creation of energy to maintain $\rho_{\Lambda}=cte$~\cite{Carroll:2000}.}
In the SSM scenario, Hoyle~\cite{Hoyle1,Hoyle2} introduced the C-field in the Einstein equations as $G_{\mu\nu}+C_{\mu\nu}=8\pi G T_{\mu\nu}$ in order to have a solution for the conundrum of the continuous creation. From a more recent perspective, this $C$-field can be expressed as $G_{\mu\nu}+\frac{1}{4}(R+8\pi G T)g_{\mu\nu}=8\pi G T_{\mu\nu}$ using unimodular gravity (UG)~\cite{Einstein,Ellis_2011,Sudarsky}, 
which contains an extra term equivalent to the C-field proposed by Hoyle but in a natural deduction. We keep in mind that UG is a model that naturally emerges from a Lagrangian, only demanding the invariance in its volume $\sqrt{-g}=\xi$, where $\xi=1$ is a constant. In this context, the vacuum energy has no direct gravitational effects, and $\Lambda$ is only an integration constant, implying that it is possible to choose a small value (or even $\Lambda=0$) as demanded in studies like~\cite{Mitras}. 

Connecting all these arguments, we raise the following proposition: {\it we are not in a position
to know whether the universe is already transiting to an accelerated or a steady-state phase using the redshift drift. (The detection of the radiation exposed hereafter could be evidence of LSS instead of an accelerated Universe. However, the radiation acts as an equivalent to CC in the standard way.)  Indeed, even if we use the luminous distance, $d_L(z)=H^{-1}z(z+1)$, or the angular distance, $d_{A}(z)=H^{-1}z(z+1)^{-1}$, for $z\lesssim0.6$, there is no experiment/observation that would allow us to conclude if the Universe is transiting to an accelerated or a late steady-state Universe.}

Thus, under the previous state of accelerated and LSS conditions, it is possible to follow Universe ~consequences. 

The traditional view for CC needs to deal with quantum vacuum fluctuations afflicted by the ultraviolet divergences that grow at $k^4$ when the energy density is calculated (see~\cite{Carroll:2000} and references therein). The reinterpretation consists in the use of a continuous creation of radiation in concordance with previous calculations for SSM, which can be deduced through~expression
\begin{eqnarray}
    \dot{\delta}(t)=-\sigma\left(\nu_1\frac{a(t_1)}{a(t)},t\right)\delta(t),
\end{eqnarray}
where $\delta(t)$ is the number of particles whose solution takes the form $\delta(t_0)=e^{-\tau}\delta(t_1)$ in terms of optical depth $\tau$, dot represents a time derivative, 
$\sigma(\nu)$ is the absorption rate of a radiation of frequency $\nu$, $a(t)$ is the scale factor related with the line element $ds^2=-dt^2+a(t)^2(dr^2+r^2d\Omega^2)$, $d\Omega^2$ is the solid angle and it is considered null curvature $k=0$ according to recent observations~\cite{Planck:2018}. Due to that, we assume that our particles follow Bose statistic; with relation $8\pi\nu^2\zeta(t,\nu)=\Gamma(\nu,t)$, we can generalize $\delta(t)$ as the following differential equation:
\begin{eqnarray}
    \frac{\dot{\delta}(t)}{\delta(t)}=-\sigma\left(\nu_1\frac{a(t_1)}{a(t)},t\right)+\zeta\left(\nu_1\frac{a(t_1)}{a(t)},t\right),
\end{eqnarray}
where $\zeta(\nu)$ is the emission rate. After some straightforward calculations, we arrive to (see details in~\cite{weinberg1972gravitation})
\begin{eqnarray}
&&n(\nu)=8\pi\nu^2\int_{t_1}^{t_0}\exp\Big\lbrace-\int_t^{t_0}\Big[\sigma\left(\nu\frac{a(t_0)}{a(t')},t'\right)\nonumber\\&&-\zeta\left(\nu\frac{a(t_0)}{a(t')},t'\right)\Big]dt'\Big\rbrace \zeta\left(\nu\frac{a(t_0)}{a(t)},t\right)dt, \label{nphot}
\end{eqnarray}
where $n(\nu)$ is the number function and $8\pi\nu^2\zeta(\nu)d\nu$ is the emission rate per unit volume of radiation 
between frequency $\nu$ and $\nu+d\nu$, with $t_0$ and $t_1$ arbitrary 
chosen. We notice that Equation~\eqref{nphot} is restricted to the scale factor, which in turn is coupled with a field theory of gravitation which in principle is GR but not restricted to possible extensions in which the scale factor is involved.

Thus, our starting point is Equation~\eqref{nphot}, and we also consider that the radiation could be standard photons but not restricted to them (i.e., other particles with similar behavior such as axions and dark photons, among others, are allowed).

In the de Sitter phase, Expression \eqref{nphot} can be written as~\cite{weinberg1972gravitation}

\begin{eqnarray}
&&n(\nu)=8\pi\nu^2\int_{-\infty}^{t_0}\exp\Big\lbrace-\int_t^{t_0}\Big[\sigma\left(\nu\exp(H_0[t_0-t'])\right)\nonumber\\&&-\zeta\left(\nu\exp(H_0[t_0-t'])\right)\Big]dt'\Big\rbrace\zeta\left(\nu\exp(H_0[t_0-t])\right)dt, \label{nphotSit}
\end{eqnarray}
where it is assumed that, during this de Sitter phase, $\dot{a}/a\approx H_0$ (this assumption is only valid for $z\leq0.6$). With an appropriate change in variables to avoid dependence on $t_0$, we~have
\begin{eqnarray}
    &&n(\nu)=8\pi\nu^2\int_{\nu}^{\infty}\frac{\zeta(\nu')}{H_0\nu'}\exp\Big(-\int_{\nu}^{\nu'}\frac{d\nu^*}{H_0\nu^*}\Big[\sigma(\nu^*)\nonumber\\&&-\zeta(\nu^*)\Big]\Big)d\nu',
\end{eqnarray}
and differentiating with respect to $\nu$, we arrive to the following expression:
\begin{equation}
\zeta(\nu)=\frac{n(\nu)\sigma(\nu)}{8\pi\nu^2+n(\nu)}+\left[2n(\nu)-\nu\frac{dn(\nu)}{d\nu}\right]\frac{H_0}{8\pi\nu^2+n(\nu)}, \label{Diffn} 
\end{equation}
where we separate terms that depend on Hubble constant $H_0$. Thus, if we stand {by} our hypothesis of assuming that the radiation (standard or dark emission) follows a Planckian number distribution, we 
 have homogeneity and isotropy
\begin{equation}
n(\nu)=8\pi \nu^2[\exp(\nu/\mathcal{T})-1]^{-1}, \label{numberdensity} 
\end{equation}
where $\mathcal{T}$ is a fixed temperature. Therefore, we obtain, from Equation~\eqref{Diffn},
\begin{eqnarray}
\zeta(\nu)=\exp(-\nu/\mathcal{T})\sigma(\nu)+\zeta(\nu)_{\rm CMV}, \label{Excess}
\end{eqnarray}
where
\begin{eqnarray}
\zeta(\nu)_{\rm CMV}=H_0\left(\frac{\nu}{\mathcal{T}}\right)[\exp(\nu/\mathcal{T})-1]^{-1}. \label{srcPlanck}
\end{eqnarray}
We notice that $\zeta(\nu)_{\rm CMV}$ is a constant emission of radiation that is independent of absorption $\sigma(\nu)$ and can be expressed as a function of the Hubble parameter, where subscript CMV indicates the Cosmic MeV emission. The first term in \eqref{Excess} (r.h.s) is the classical behavior emission--absorption term for radiation implying interactions with baryonic matter and astrophysical background lights, such as the extragalactic background light (EBL) and the CMB. \footnote{This interaction depends on the particle we are dealing with; for example, in the dark sector, interactions are weaker than those associated with standard photons.}

We also notice that even in the case $\nu\to0$, we always have a non-negligible emission in the form $\zeta(0)_{\rm CMV}=H_0$. After an integration in the form of $\rho=8\pi\int_0^{\infty}\nu^2\zeta(\nu)_{\rm CMV}d\nu$ (which is convergent) for energy density, we conclude
\begin{equation}
\rho_{\rm CMV}=\frac{\pi^2}{15} H_0\mathcal{T}^3.\label{New}
\end{equation}

Due to the equivalence between an accelerated and a stationary Universe, it is possible to propose identification $\rho_{\rm CMV}=\rho_{\Lambda}\simeq2.46\times10^{-11}{\rm eV}^4$, where the last number is the value expected for the cause of Universe acceleration. Consequently, we arrive to a fluid with energy 
\begin{equation}\label{eq:30MeV}
E_{\rm CMV}\simeq29.5\, {\rm MeV}.  
\end{equation}
The energy is obtained by using the current value for the Hubble constant reported in~\cite{Planck:2019nip} 
($E_{\rm CMV}\simeq28.8\,$MeV  using a local measurement of $H_0$~\cite{Riess:2020fzl}). This energy region for standard photons is a minimum in the photon interaction cross-section, and the transition from Compton scattering to pair production as the dominant process makes a piece of new physics evidence in this region particularly challenging. This would explain why it has not been detected yet. Consequently, in the following sections, we investigate the possibility of detection through astrophysical tests.

\section{Astrophysical Tests}\label{Sec:AstroTests}

In a CMB-like scenario,  Universe acceleration forecasts an isotropic and homogeneous cosmic background of standard light radiation, such as the known CMB~\cite{PenziasAndWilson,Planck:2018}. 
Thus, if $E_{\rm CMV}$ is a CMB-like radiation, it follows the energy density distribution given by Equation~(\ref{numberdensity}) as black-body radiation with a mean energy density given by Equation~(\ref{eq:30MeV}). As a background light, it is likely to interact with gamma rays and annihilate by the photon pair-production process ($\gamma~\gamma~\rightarrow~e^{+}~e^{-}$). The astrophysical expected outcome of this process is attenuating the expected astrophysical photon flux; for a CMB energy range, such attenuation is expected on the ultra-high-energy gamma rays. Lighter background photons as the EBL attenuates the TeV gamma-ray flux of extragalactic point sources, such as blazars~\cite{DeAngelis,Gilmore:2011ks,Franceschini:2008tp,Dominguez:2010bv}.
In this line of thought, we find the survivable probability ($P_{surv}=e^{-\tau}$) of standard photons, such as gamma rays, by finding the optical depth ($\tau$) but considering the presence of the $E_{\rm CMV}$ radiation as a background light. The survivable probability is given by 
{
\begin{align}
P_{surv}=  \exp & \left[ -\int_0^z  \frac{{\rm cd}z}{H_0 (1+z) h(z)} \int_{-1}^1 {\rm d} \cos \theta \frac{1 - \cos \theta}{2} \right.
\\ & \left.  \times \int_{\epsilon_{th}}^{\infty} {\rm d}\epsilon n(\epsilon, z) \nonumber \sigma(E_\gamma, \epsilon, z)\right],
\end{align}
}where $H_0$ stands for the Hubble constant in the present time, $h(z)=\sqrt{\Omega_{m}(1+z)^{3} + \Omega_{\rm CMV}}$ is the distance element in an expanding universe, $\sigma (E_\gamma, \epsilon, z)$ is the Breit--Wheeler cross-section for pair production process $\gamma~\gamma~\rightarrow~e^{+}~e^{-}$~\cite{Breit:1934zz}, and $n(\epsilon,z)$ is the background density given by Equation~\eqref{numberdensity}, with $\epsilon=\nu$.

\begin{figure*}
    \centering
    \includegraphics[width=0.7\textwidth]{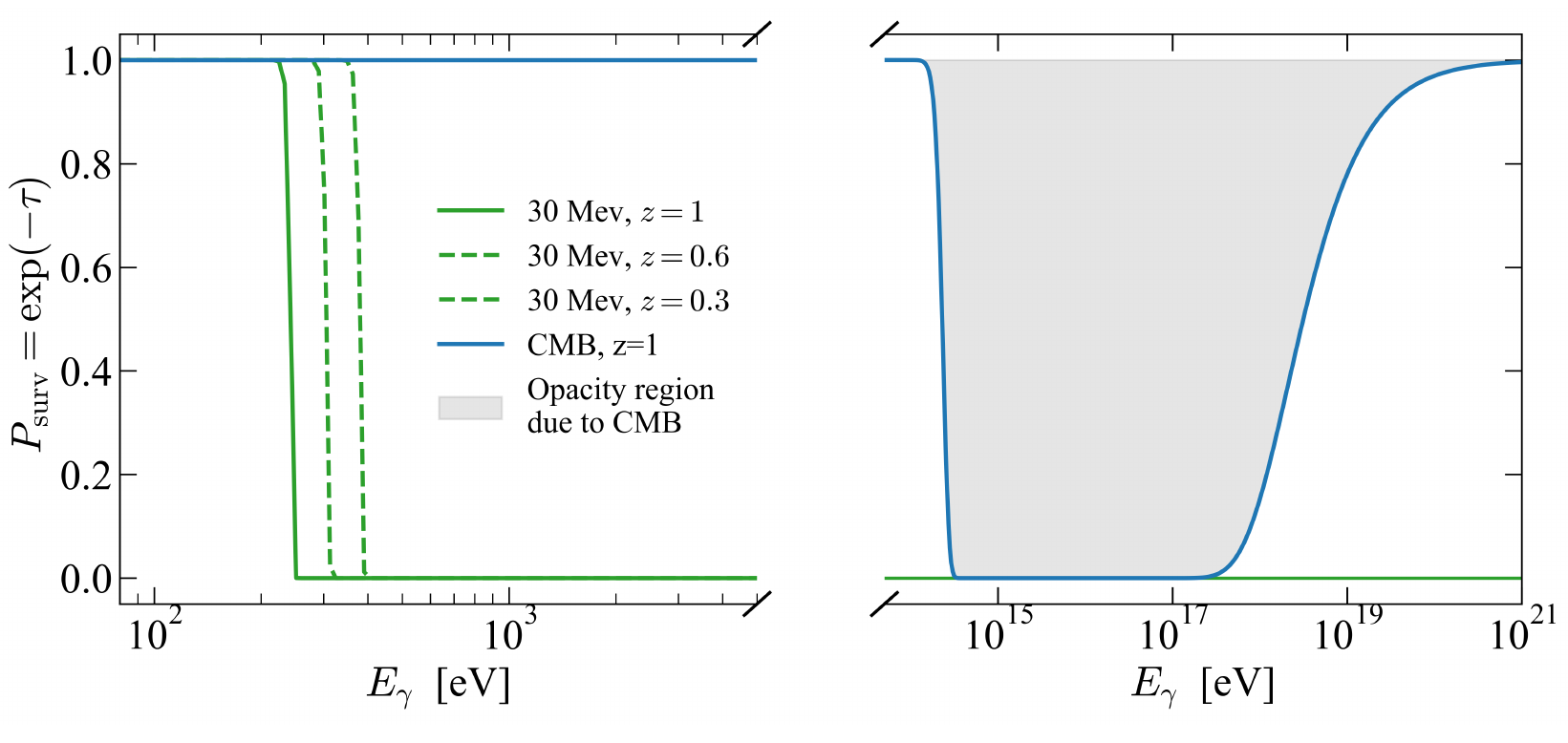}
    \caption{Photon survival probability due to interaction with the 30 MeV radiation from a given z as standard photons. For comparison, survival probability due to CMB is included. Everything above such lines is attenuated; we illustrate this as the shadow area in the CMB case.}
    \label{fig:Prob}
\end{figure*}

The outcome of these is that if $E_{CMV}$ radiation is CMB-like, any astrophysical photon above the keV energy range and from a distance beyond $~10^{-26}$~Mpc will be attenuated by the presence of $E_{CMV}$ emission. 
In Figure~\ref{fig:Prob}, we show the survival probability of gamma rays from different $z$'s to us on Earth. For comparison, we include the gamma-ray survival probability due to CMB; the shadow area represents the attenuated region for gamma rays from $z=1$, from which CMB allows ultra-high-energy photon propagation from $z=1$ to Earth. Once again, we conclude that non-standard photons of the keV order would survive $E_{CMV}$ radiation, which contradicts the observations. Therefore, if CMV emission exists, it cannot be made by standard photons.

Thus, considering the unknown nature of the CMV and assuming that it can decay to some standard photons, it may be proven through astrophysical tests.
Experiments like the Compton Telescope (COMPTEL)~\cite{COMPTEL01} and the Energetic Gamma-ray Experiment Telescope (EGRET)~\cite{EGRET01} have reported measurements of photon flux at the energy region of our interest~\cite{Ajello_2008}. 

\begin{figure*}
    \centering
    \includegraphics[width=0.55\textwidth]{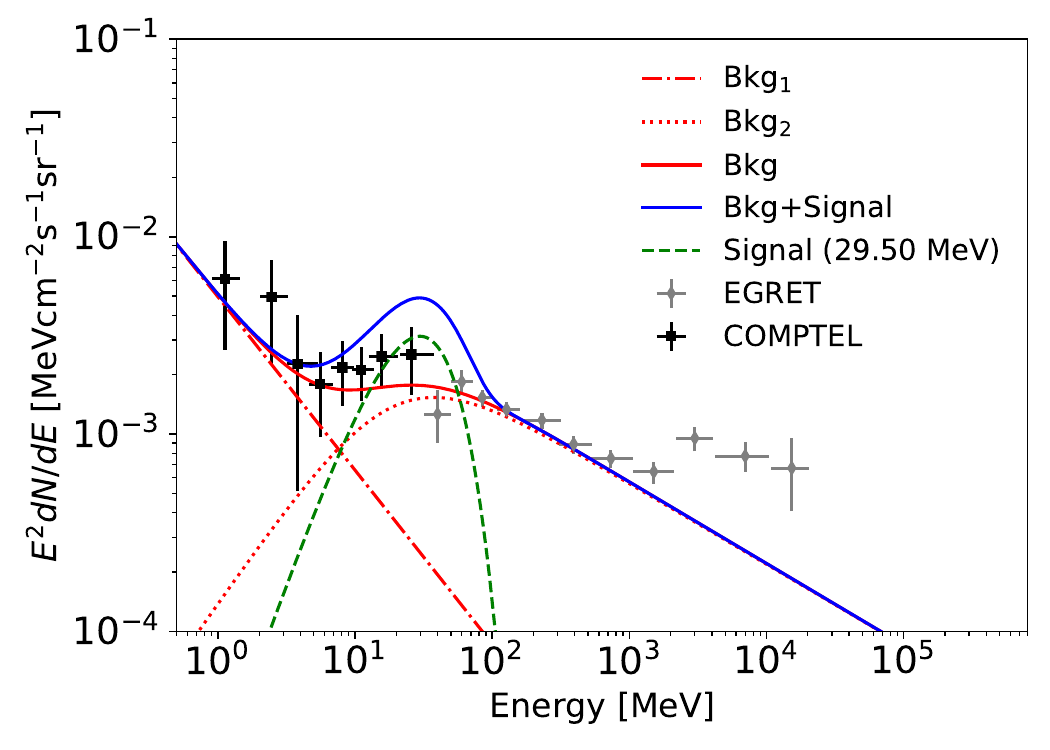}
    \caption{Energy flux vs energy. Black and grey markers are COMPTEL and EGRET data, respectively. The solid blue line corresponds to Bkg+Signal fit, and the solid red line to Bkg is composed of Bkg$_1$ (dot-dashed red line) and Bkg$_2$ (dotted red line).}
    \label{fig:fit}
\end{figure*}

Therefore, in order to test the predicted signal, we model EGRET and COMPTEL data extracted from~\cite{COMPTEL01,EGRET01} of the energy flux in the energy range $1.12<E<15.28 \times 10^3 {\rm }$ MeV. Based on~\cite{Oberlack:2010}, we consider the background modeled by the sum of two components, ${\rm Bkg}_1(E) + {\rm Bkg}_2(E)$, where  
\begin{equation}
    {\rm Bkg}_{1,2}(E) = \frac{C_{1,2}}{(E/E_b)^{\Gamma_{1,3}}+ (E/E_b)^{\Gamma_{2,4}}},
\end{equation}
where $C_{1,2}$ are normalization constants, $E_b$ is the energy peak, and $\Gamma_1$ and $\Gamma_2$ are constants. For the ${\rm Bkg}_1$ component, we fix $\Gamma_1=1.32$, $\Gamma_2=2.88$, $E_b=25\, {\rm keV}$, and for the ${\rm Bkg}_2$ component, $\Gamma_3=1.0$, $\Gamma_4=2.41$ and $E_b=20\, {\rm MeV}$~\cite{Oberlack:2010}, and we allow variation of both $C_1$ and $C_2$. The full model is obtained by adding the signal presented in Equation~\eqref{srcPlanck}, which is named 
Bkg+Signal. Figure \ref{fig:fit} shows the fit obtained, the solid blue line corresponds to Signal+Bkg and the solid red line is the total background.

We compare the two models, Bkg and Bkg+Signal, statistically through the Akaike information criterion corrected (AICc)~\cite{Akaike:1974,Sugiura:1978} for small samples defined as ${\rm AICc} = \chi^2_{min} + 2k + (2k^2+2k)/(N-k-1)$ where $\chi^2_{min}$ is the minimum of the $\chi^2$-function, $k$ is the number of degrees of freedom and $N$ is the size of the sample. In this criterion, the model with the lower value of AICc is the one preferred by data. When difference $\Delta{\rm AICc}$ between a given model and the best one is $\Delta{\rm AICc}<4$,
both models are equally supported by the data. For the range $4<\Delta{\rm AICc}<10$, the data still support the given model but less than the preferred one, and if $\Delta{\rm AICc}>10$, the given model is not supported. Our result gives $\Delta {\rm AICc}={\rm AICc(Bkg+Signal)}-{\rm AICc(Bkg)}= -118.8$, which indicates that the model Bkg+Signal is preferred by EGRET and COMPTEL datasets. Additionally, we find signal amplitude $A= 0.151\pm 0.014 \,$MeVcm$^{-2}$s$^{-1}$sr$^{-1}$.

As can be seen, the expected energy is in the extreme of both data sets, in the maximum of COMPTEL and in the minimum of EGRET, causing difficulties in its detection through the recent data compilations (see Figure \ref{fig:fit}).
Future experiments could improve the signal-to-noise detection in the MeV region, such as the GRAMS Project~\cite{ARAMAKI2020107} and GammaTPC~\cite{GammaTPC}. They might be able to confirm whether CMV emission exists, and if it does, the cause of the current de Sitter transition might be explained. 

\section{The Underneath Field Theory, a Revision} \label{Sec:FieldTheory} 

The continuous creation of energy, in principle, is incompatible with GR because $\nabla^{\mu}T_{\mu\nu}\neq0$ unless we accept fluids with the $w<-1/3$ equation of state. One of the best approaches under a field equation to have a non-conservation of the energy--momentum tensor and compatible with the fluid discussed in Section~\ref{LSS} is through the equations of unimodular gravity (UG) given by 

\begin{equation}
R_{\mu\nu}-\frac{1}{4}g_{\mu\nu}R=8\pi G\left(T_{\mu\nu}-\frac{1}{4}g_{\mu\nu}T\right),
\end{equation}
which clearly is traceless~\cite{Einstein}, $R_{\mu\nu}$, $R$ are the Ricci tensor and scalar, respectively, $T_{\mu\nu}$ is the energy--momentum tensor, $T=g^{\mu\nu}T_{\mu\nu}$ is the energy--momentum scalar and $G$ is the Newton gravitational constant. In this context, a CC is unnecessary, and traditional fluids can produce evolution $a(t)=\exp\Lambda(t-t_0)$ only under the restriction that $\dot{\rho}^{fluid}=0$, caused naturally by 

\begin{equation}
32\pi G\nabla^{\mu}T_{\mu\nu}=\nabla^{\mu}(R+8\pi GT)g_{\mu\nu}, \label{noncons}
\end{equation}
where the general covariance is not demanded. We notice that this approach is similar to the stationary model~\cite{BondiandGold,Hoyle1,Hoyle2} in which a continuous creation of matter/energy produces a de Sitter behavior, indistinguishable from the scale factor for an accelerated Universe. Some studies~\cite{Garcia-Aspeitia:2019yni,Garcia-Aspeitia:2019yod} suggest that a continuous creation of radiation (relativistic particles, axions, dark photons) under the UG approach could resolve the problem of the observed de Sitter phase, without the necessity of $w<-1/3$, demanded in standard GR.

Assuming a Friedmann--Lemaitre--Robertson--Walker (FLRW) metric and a perfect fluid energy--momentum tensor, we arrive at the following equation:
\begin{equation}
    \dot{H}=-4\pi G\sum_i(\rho_i+p_i), \label{Hdot}
\end{equation}
where $\rho_i$, $p_i$ are the density and pressure of the fluids, respectively, $H\equiv\dot{a}/a$ is the Hubble parameter, $a$ is the scale factor, and dot stands for time derivative. On the other hand, resolving for Equation~\eqref{noncons}, we have 
\begin{equation}
    \sum_i[(\dot{\rho}_i+\dot{p}_i)+3H(\rho_i+p_i)]=\frac{H^3}{4\pi G}(1-j), \label{j}
\end{equation}
where $j\equiv\dddot{a}/aH^3$ is known as the jerk parameter~\cite{Zhang:2016}  where the dots stand for third-order derivatives. The third-order derivatives that this theory contains are tamed through the cosmographic jerk parameter, which also acts as a source term. Thus, after some manipulation using Equations~\eqref{Hdot} and \eqref{j}, the dynamical equations for the cosmology in this context can be presented as follows:
\begin{eqnarray}
    &&H^2=\frac{8\pi G}{3}\sum_i\rho_i+H^2_{UG}, \\
    &&\frac{\ddot{a}}{a}=-\frac{4\pi G}{3}\sum_i(\rho_i+3p_i)+H^2_{UG}, \\
    &&H^2_{UG}=\frac{8\pi G}{3}\sum_ip_i+\frac{2}{3}\int_{a_{ini}}^aH(a')^2[j(a')-1]\frac{da'}{a'},
\end{eqnarray}
where subscript UG refers to Unimodular Gravity.  According  to~\cite{Garcia-Aspeitia:2019yni,Garcia-Aspeitia:2019yod}, the best selection of $j$, to mimic the $\Lambda$CDM model but taking into account the clues about the causative of the Universe's acceleration, is $j=\frac{9}{2}(1+w)wE(z)^{-2}\Omega_{0i}(z+1)^{3(w+1)}+1$, where $w$ is the fluid equation of state and $\Omega_{0i}$ is the density parameter, $z=a^{-1}-1$ is the redshift and $E(z)\equiv H(z)/H_0$. The election, naturally, allows us to decouple matter with the standard equation of state while radiation depends on the $j$ parameter. Therefore, radiation plays a central role in the Universe's acceleration due to the coupling with $j$. Thus, demanding a continuity equation for new dark energy term gives us the following Friedmann equation

\begin{equation}
    E(z)^2=\Omega_{0m}(z+1)^3+\Omega_{0r}(z+1)^4+\omega_{\rm CMV}\Omega_{0{\rm CMV}}(z_{ini}+1)^4, \label{Friedm}
\end{equation}
where the last term acts like a cosmological constant but with an origin based in the term $\frac{1}{4}(R+8\pi GT)g_{\mu\nu}$. This constant acts like a diffusion parameter and can be interpreted as a CMV component plus the effects of the UG dictated by the $z_{ini}$ free parameter (see~\cite{Garcia-Aspeitia:2019yni,Garcia-Aspeitia:2019yod} for details), where $\Omega_{0m}$ and $\Omega_{0{\rm CMV}}$ are the matter and CMV parameters, respectively, $\omega_{\rm CMV}=1/3$ and $z_{ini}=11.473^{+0.074}_{-0.073}$ constrained through recent observations (see~\cite{Garcia-Aspeitia:2019yni,Garcia-Aspeitia:2019yod}) and interpreted as the region where the term that generates the de Sitter expansion emerges (but not dominates).

\section{Discussion and Conclusions} \label{DC}

As we discussed previously, at the moment, it is impossible to differentiate between an accelerated or late stationary Universe because the observables ($d_L$ and $d_A$) in terms of the redshift drift of all the experiments that measure the Universe evolution are equally compatible with both approaches. 
We emphasize that this argument is only valid for $z\lesssim0.6$, where the data confirms the transition.

On the other hand, it is well known that both models (accelerated and stationary) need a continuous creation of energy/matter to maintain the energy density constant of some species in the Universe,  which, in turn, pushes the GR's limits in its current form and suggests that changes are required to obtain, for instance, a non-conservation to the energy--momentum tensor.
This equivalence suggests a new strategy to simultaneously tackle the energy density problem of the CC and the interpretation of the emission in the regime of $\sim$29.5 MeV, which, in principle, could be detected through decaying to standard photons. {\it Confirming this signal is crucial, so we encourage further astrophysical analysis in this scope.} Suggestively, the MeV region is also connected to several new physics proposals, such as Primordial Black Holes (PBH)~\cite{Zeldovich1966,Hawking1971,Carr_Hawking:1974} emitting radiation through Hawking effect, self-annihilation of DM particles candidates with MeV masses that can produce gamma radiation~\cite{Gonzalez-Morales:2017jkx,Boudaud:2018oya}, and some quantum gravity effects (like Continuous Spontaneous Localization, compatible with UG~\cite{Sudarsky} or the Diosi--Penrose model, DP~\cite{Diosi1,Diosi2,Penrose}) could also produce radiation emission at the same energy scale of tens of MeV. 
Identifying the event that causes the excess radiation is vital because the theory presented here cannot identify the event that causes the previously mentioned excess. For example, PBH as DM implies decaying radiation at MeV generating the observed accelerated/stationary Universe, sustaining a unified framework that relates several events. In this case, the excess of radiation is a transitory effect; therefore, the accelerated/stationary Universe is also a transitory process. For example, in the case where the cause of the excess of radiation is the DP model under charged particles, the collapse of the wave function by gravitational effects has significant consequences for the evolution of the Universe and its current stationary/accelerated stage. The radiation energy for this case is in the range of $\Delta E=(10-10^5)$ keV, with CMV inside the expected energy region. Other particles like axions~\cite{Axions} or dark photons~\cite{DarkPhotons} could be possible candidates. Therefore, models could prove that we are dealing with an LSS instead of an accelerated Universe.

We need to remark that a stationary Universe is not a natural state (at least under the Friedmann--Lemaitre--Robertson--Walker line element) because maintaining constant energy density requires the violation of the conservation of the energy--momentum tensor. As a consequence, we expect that this state will eventually stop, finishing this particular condition. However, to respond to the question of how to differentiate between the two universes, we need to know what the source of the CMV radiation is, if it exists. In this work, we showed that the survival probability of standard photons due to the CMV predicts an impossible opacity region, not allowing the detection of any $>$keV astrophysical photons on Earth. This strongly constrains the hypothesis that this radiation exists as standard photons. Hence, we also studied the possibility of an indirect signal of standard photons derived from a potential non-standard component of the CMV emission at 29.5 MeV. Our findings suggest that the model Bkg+Signal is preferred over the only Bkg scenario, using EGRET and COMPTEL datasets. 

Finally, we must remember that a change in variables to the de Sitter line element produces a stationary metric as $ds^2=-(1-r^2/\alpha^2)dt^2+(1-r^2/\alpha^2)^{-1}dr^2+r^2d\Omega^2$, where $\alpha$ is a nonzero constant of the hyperboloid of one sheet. Remarkably, a de Sitter evolution is analogous to a stationary Universe, as discussed throughout this paper. Additionally, we emphasize that when using standard GR, a fluid with $w<-1/3$ EoS is necessary in order to have a de Sitter Universe. 
We also do not have the physics of DE, and thus it is impossible to know whether the particles decay in an energy region in order to be detected by some experiment. Under these conditions, it is not possible to have the same predictions presented in this paper. 

We encourage further and novel astrophysical experiments and studies to unravel the $29.5$ MeV emission in order to fully comprehend the current de Sitter stage of our Universe. 

\begin{acknowledgements}
YBAP acknowledges the Ph.D. grant provided by CONACYT. MAGA acknowledges support from c\'atedra Marcos Moshinsky and Universidad Iberoamericana for the support with the National Research System (SNI) grant. We thank the anonymous referees for their thoughtful remarks and suggestions. The numerical analysis was carried out by {\it Numerical Integration for Cosmological Theory and Experiments in High-energy Astrophysics} (Nicte Ha) cluster at IBERO University, acquired through c\'atedra MM support.  A.H.A. thank for the support from Luis Aguilar, Alejandro de Le\'on, Carlos Flores, and Jair Garc\'ia of the Laboratorio Nacional de Visualizaci\'on Cient\'ifica Avanzada. 
A.H.A. and M.A.G.-A. acknowledge partial {support} from project ANID Vinculaci\'on Internacional FOVI220144. We are also grateful to Ver\'onica Motta for revising this paper, suggesting changes, and correcting misinterpretations in the text. This paper is in memory of my father Miguel \'Angel. 
\end{acknowledgements}


\begin{thebibliography}{48}%
\makeatletter
\providecommand \@ifxundefined [1]{%
 \@ifx{#1\undefined}
}%
\providecommand \@ifnum [1]{%
 \ifnum #1\expandafter \@firstoftwo
 \else \expandafter \@secondoftwo
 \fi
}%
\providecommand \@ifx [1]{%
 \ifx #1\expandafter \@firstoftwo
 \else \expandafter \@secondoftwo
 \fi
}%
\providecommand \natexlab [1]{#1}%
\providecommand \enquote  [1]{``#1''}%
\providecommand \bibnamefont  [1]{#1}%
\providecommand \bibfnamefont [1]{#1}%
\providecommand \citenamefont [1]{#1}%
\providecommand \href@noop [0]{\@secondoftwo}%
\providecommand \href [0]{\begingroup \@sanitize@url \@href}%
\providecommand \@href[1]{\@@startlink{#1}\@@href}%
\providecommand \@@href[1]{\endgroup#1\@@endlink}%
\providecommand \@sanitize@url [0]{\catcode `\\12\catcode `\$12\catcode
  `\&12\catcode `\#12\catcode `\^12\catcode `\_12\catcode `\%12\relax}%
\providecommand \@@startlink[1]{}%
\providecommand \@@endlink[0]{}%
\providecommand \url  [0]{\begingroup\@sanitize@url \@url }%
\providecommand \@url [1]{\endgroup\@href {#1}{\urlprefix }}%
\providecommand \urlprefix  [0]{URL }%
\providecommand \Eprint [0]{\href }%
\providecommand \doibase [0]{http://dx.doi.org/}%
\providecommand \selectlanguage [0]{\@gobble}%
\providecommand \bibinfo  [0]{\@secondoftwo}%
\providecommand \bibfield  [0]{\@secondoftwo}%
\providecommand \translation [1]{[#1]}%
\providecommand \BibitemOpen [0]{}%
\providecommand \bibitemStop [0]{}%
\providecommand \bibitemNoStop [0]{.\EOS\space}%
\providecommand \EOS [0]{\spacefactor3000\relax}%
\providecommand \BibitemShut  [1]{\csname bibitem#1\endcsname}%
\let\auto@bib@innerbib\@empty
\bibitem [{\citenamefont {Riess}\ \emph {et~al.}(1998)\citenamefont {Riess},
  \citenamefont {Filippenko}, \citenamefont {Challis}, \citenamefont
  {Clocchiatti}, \citenamefont {Diercks} \emph {et~al.}}]{Riess:1998}%
  \BibitemOpen
  \bibfield  {author} {\bibinfo {author} {\bibfnamefont {A.~G.}\ \bibnamefont
  {Riess}}, \bibinfo {author} {\bibfnamefont {A.~V.}\ \bibnamefont
  {Filippenko}}, \bibinfo {author} {\bibfnamefont {P.}~\bibnamefont {Challis}},
  \bibinfo {author} {\bibfnamefont {A.}~\bibnamefont {Clocchiatti}}, \bibinfo
  {author} {\bibfnamefont {A.}~\bibnamefont {Diercks}},  \emph {et~al.},\
  }\href {http://stacks.iop.org/1538-3881/116/i=3/a=1009} {\bibfield  {journal}
  {\bibinfo  {journal} {The Astronomical Journal}\ }\textbf {\bibinfo {volume}
  {116}},\ \bibinfo {pages} {1009} (\bibinfo {year} {1998})}\BibitemShut
  {NoStop}%
\bibitem [{\citenamefont {Perlmutter}\ \emph {et~al.}(1999)\citenamefont
  {Perlmutter}, \citenamefont {Aldering}, \citenamefont {Goldhaber},
  \citenamefont {Knop}, \citenamefont {Nugent}, \citenamefont {others},\ and\
  \citenamefont {Project}}]{Perlmutter:1999}%
  \BibitemOpen
  \bibfield  {author} {\bibinfo {author} {\bibfnamefont {S.}~\bibnamefont
  {Perlmutter}}, \bibinfo {author} {\bibfnamefont {G.}~\bibnamefont
  {Aldering}}, \bibinfo {author} {\bibfnamefont {G.}~\bibnamefont {Goldhaber}},
  \bibinfo {author} {\bibfnamefont {R.~A.}\ \bibnamefont {Knop}}, \bibinfo
  {author} {\bibfnamefont {P.}~\bibnamefont {Nugent}}, \bibinfo {author}
  {\bibnamefont {others}}, \ and\ \bibinfo {author} {\bibfnamefont {T.~S.~C.}\
  \bibnamefont {Project}},\ }\href
  {http://stacks.iop.org/0004-637X/517/i=2/a=565} {\bibfield  {journal}
  {\bibinfo  {journal} {The Astrophysical Journal}\ }\textbf {\bibinfo {volume}
  {517}},\ \bibinfo {pages} {565} (\bibinfo {year} {1999})}\BibitemShut
  {NoStop}%
\bibitem [{\citenamefont {Aghanim}\ \emph
  {et~al.}(2020{\natexlab{a}})\citenamefont {Aghanim} \emph
  {et~al.}}]{Planck:2018}%
  \BibitemOpen
  \bibfield  {author} {\bibinfo {author} {\bibfnamefont {N.}~\bibnamefont
  {Aghanim}} \emph {et~al.},\ }\href {\doibase 10.1051/0004-6361/201833910}
  {\bibfield  {journal} {\bibinfo  {journal} {Astronomy {\&} Astrophysics}\
  }\textbf {\bibinfo {volume} {641}},\ \bibinfo {pages} {A6} (\bibinfo {year}
  {2020}{\natexlab{a}})}\BibitemShut {NoStop}%
\bibitem [{\citenamefont {Zeldovich}(1968)}]{Zeldovich}%
  \BibitemOpen
  \bibfield  {author} {\bibinfo {author} {\bibfnamefont {Y.~B.}\ \bibnamefont
  {Zeldovich}},\ }\href@noop {} {\bibfield  {journal} {\bibinfo  {journal}
  {Soviet Physics Uspekhi}\ }\textbf {\bibinfo {volume} {11}} (\bibinfo {year}
  {1968})}\BibitemShut {NoStop}%
\bibitem [{\citenamefont {Weinberg}(1989)}]{Weinberg}%
  \BibitemOpen
  \bibfield  {author} {\bibinfo {author} {\bibfnamefont {S.}~\bibnamefont
  {Weinberg}},\ }\href@noop {} {\bibfield  {journal} {\bibinfo  {journal}
  {Reviews of Modern Physics}\ }\textbf {\bibinfo {volume} {61}} (\bibinfo
  {year} {1989})}\BibitemShut {NoStop}%
\bibitem [{\citenamefont {Bondi}\ and\ \citenamefont
  {Gold}(1948)}]{BondiandGold}%
  \BibitemOpen
  \bibfield  {author} {\bibinfo {author} {\bibfnamefont {H.}~\bibnamefont
  {Bondi}}\ and\ \bibinfo {author} {\bibfnamefont {T.}~\bibnamefont {Gold}},\
  }\href {\doibase 10.1093/mnras/108.3.252} {\bibfield  {journal} {\bibinfo
  {journal} {Monthly Notices of the Royal Astronomical Society}\ }\textbf
  {\bibinfo {volume} {108}},\ \bibinfo {pages} {252} (\bibinfo {year}
  {1948})},\ \Eprint
  {http://arxiv.org/abs/https://academic.oup.com/mnras/article-pdf/108/3/252/8076977/mnras108-0252.pdf}
  {https://academic.oup.com/mnras/article-pdf/108/3/252/8076977/mnras108-0252.pdf}
  \BibitemShut {NoStop}%
\bibitem [{\citenamefont {Weinberg}(1972)}]{weinberg1972gravitation}%
  \BibitemOpen
  \bibfield  {author} {\bibinfo {author} {\bibfnamefont {S.}~\bibnamefont
  {Weinberg}},\ }\href {https://books.google.com.mx/books?id=XLbvAAAAMAAJ}
  {\emph {\bibinfo {title} {Gravitation and Cosmology.}}}\ (\bibinfo
  {publisher} {Wiley},\ \bibinfo {year} {1972})\BibitemShut {NoStop}%
\bibitem [{\citenamefont {Trevisani}\ and\ \citenamefont
  {Lima}(2023)}]{Trevisani:2023wpw}%
  \BibitemOpen
  \bibfield  {author} {\bibinfo {author} {\bibfnamefont {S.~R.~G.}\
  \bibnamefont {Trevisani}}\ and\ \bibinfo {author} {\bibfnamefont {J.~A.~S.}\
  \bibnamefont {Lima}},\ }\href {\doibase 10.1140/epjc/s10052-023-11301-8}
  {\bibfield  {journal} {\bibinfo  {journal} {The European Physical Journal C}\
  }\textbf {\bibinfo {volume} {83}} (\bibinfo {year} {2023}),\
  10.1140/epjc/s10052-023-11301-8}\BibitemShut {NoStop}%
\bibitem [{\citenamefont {C\'ardenas}\ \emph {et~al.}(2020)\citenamefont
  {C\'ardenas}, \citenamefont {Cruz},\ and\ \citenamefont
  {Lepe}}]{Cardenas:2020exv}%
  \BibitemOpen
  \bibfield  {author} {\bibinfo {author} {\bibfnamefont {V.~H.}\ \bibnamefont
  {C\'ardenas}}, \bibinfo {author} {\bibfnamefont {M.}~\bibnamefont {Cruz}}, \
  and\ \bibinfo {author} {\bibfnamefont {S.}~\bibnamefont {Lepe}},\ }\href
  {\doibase 10.1103/PhysRevD.102.123543} {\bibfield  {journal} {\bibinfo
  {journal} {Phys. Rev. D}\ }\textbf {\bibinfo {volume} {102}},\ \bibinfo
  {pages} {123543} (\bibinfo {year} {2020})},\ \Eprint
  {http://arxiv.org/abs/2008.12403} {arXiv:2008.12403 [gr-qc]} \BibitemShut
  {NoStop}%
\bibitem [{\citenamefont {Lima}\ \emph {et~al.}(2016)\citenamefont {Lima},
  \citenamefont {Santos},\ and\ \citenamefont {Cunha}}]{Lima:2015xpa}%
  \BibitemOpen
  \bibfield  {author} {\bibinfo {author} {\bibfnamefont {J.~A.~S.}\
  \bibnamefont {Lima}}, \bibinfo {author} {\bibfnamefont {R.~C.}\ \bibnamefont
  {Santos}}, \ and\ \bibinfo {author} {\bibfnamefont {J.~V.}\ \bibnamefont
  {Cunha}},\ }\href {\doibase 10.1088/1475-7516/2016/03/027} {\bibfield
  {journal} {\bibinfo  {journal} {JCAP}\ }\textbf {\bibinfo {volume} {03}},\
  \bibinfo {pages} {027} (\bibinfo {year} {2016})},\ \Eprint
  {http://arxiv.org/abs/1508.07263} {arXiv:1508.07263 [gr-qc]} \BibitemShut
  {NoStop}%
\bibitem [{\citenamefont {Vargas~dos Santos}\ \emph {et~al.}(2014)\citenamefont
  {Vargas~dos Santos}, \citenamefont {Waga},\ and\ \citenamefont
  {Ramos}}]{VargasdosSantos:2014ddg}%
  \BibitemOpen
  \bibfield  {author} {\bibinfo {author} {\bibfnamefont {M.}~\bibnamefont
  {Vargas~dos Santos}}, \bibinfo {author} {\bibfnamefont {I.}~\bibnamefont
  {Waga}}, \ and\ \bibinfo {author} {\bibfnamefont {R.~O.}\ \bibnamefont
  {Ramos}},\ }\href {\doibase 10.1103/PhysRevD.90.127301} {\bibfield  {journal}
  {\bibinfo  {journal} {Phys. Rev. D}\ }\textbf {\bibinfo {volume} {90}},\
  \bibinfo {pages} {127301} (\bibinfo {year} {2014})},\ \Eprint
  {http://arxiv.org/abs/1412.5203} {arXiv:1412.5203 [astro-ph.CO]} \BibitemShut
  {NoStop}%
\bibitem [{\citenamefont {Ramos}\ \emph {et~al.}(2014)\citenamefont {Ramos},
  \citenamefont {Vargas~dos Santos},\ and\ \citenamefont
  {Waga}}]{Ramos:2014dba}%
  \BibitemOpen
  \bibfield  {author} {\bibinfo {author} {\bibfnamefont {R.~O.}\ \bibnamefont
  {Ramos}}, \bibinfo {author} {\bibfnamefont {M.}~\bibnamefont {Vargas~dos
  Santos}}, \ and\ \bibinfo {author} {\bibfnamefont {I.}~\bibnamefont {Waga}},\
  }\href {\doibase 10.1103/PhysRevD.89.083524} {\bibfield  {journal} {\bibinfo
  {journal} {Phys. Rev. D}\ }\textbf {\bibinfo {volume} {89}},\ \bibinfo
  {pages} {083524} (\bibinfo {year} {2014})},\ \Eprint
  {http://arxiv.org/abs/1404.2604} {arXiv:1404.2604 [astro-ph.CO]} \BibitemShut
  {NoStop}%
\bibitem [{\citenamefont {Carroll}(2001)}]{Carroll:2000}%
  \BibitemOpen
  \bibfield  {author} {\bibinfo {author} {\bibfnamefont {S.~M.}\ \bibnamefont
  {Carroll}},\ }\href {\doibase 10.12942/lrr-2001-1} {\bibfield  {journal}
  {\bibinfo  {journal} {Living Rev. Rel.}\ }\textbf {\bibinfo {volume} {4}},\
  \bibinfo {pages} {1} (\bibinfo {year} {2001})},\ \Eprint
  {http://arxiv.org/abs/astro-ph/0004075} {arXiv:astro-ph/0004075 [astro-ph]}
  \BibitemShut {NoStop}%
\bibitem [{\citenamefont {Hoyle}(1948)}]{Hoyle1}%
  \BibitemOpen
  \bibfield  {author} {\bibinfo {author} {\bibfnamefont {F.}~\bibnamefont
  {Hoyle}},\ }\href {\doibase 10.1093/mnras/108.5.372} {\bibfield  {journal}
  {\bibinfo  {journal} {Monthly Notices of the Royal Astronomical Society}\
  }\textbf {\bibinfo {volume} {108}},\ \bibinfo {pages} {372} (\bibinfo {year}
  {1948})},\ \Eprint
  {http://arxiv.org/abs/https://academic.oup.com/mnras/article-pdf/108/5/372/8073757/mnras108-0372.pdf}
  {https://academic.oup.com/mnras/article-pdf/108/5/372/8073757/mnras108-0372.pdf}
  \BibitemShut {NoStop}%
\bibitem [{\citenamefont {Hoyle}(1949)}]{Hoyle2}%
  \BibitemOpen
  \bibfield  {author} {\bibinfo {author} {\bibfnamefont {F.}~\bibnamefont
  {Hoyle}},\ }\href {\doibase 10.1093/mnras/109.3.365} {\bibfield  {journal}
  {\bibinfo  {journal} {Monthly Notices of the Royal Astronomical Society}\
  }\textbf {\bibinfo {volume} {109}},\ \bibinfo {pages} {365} (\bibinfo {year}
  {1949})},\ \Eprint
  {http://arxiv.org/abs/https://academic.oup.com/mnras/article-pdf/109/3/365/8073909/mnras109-0365.pdf}
  {https://academic.oup.com/mnras/article-pdf/109/3/365/8073909/mnras109-0365.pdf}
  \BibitemShut {NoStop}%
\bibitem [{\citenamefont {Einstein}(1919)}]{Einstein}%
  \BibitemOpen
  \bibfield  {author} {\bibinfo {author} {\bibfnamefont {A.}~\bibnamefont
  {Einstein}},\ }\href@noop {} {\bibfield  {journal} {\bibinfo  {journal} {Siz.
  Preuss. Acad. Scis.}\ } (\bibinfo {year} {1919})}\BibitemShut {NoStop}%
\bibitem [{\citenamefont {Ellis}\ \emph {et~al.}(2011)\citenamefont {Ellis},
  \citenamefont {van Elst}, \citenamefont {Murugan},\ and\ \citenamefont
  {Uzan}}]{Ellis_2011}%
  \BibitemOpen
  \bibfield  {author} {\bibinfo {author} {\bibfnamefont {G.~F.~R.}\
  \bibnamefont {Ellis}}, \bibinfo {author} {\bibfnamefont {H.}~\bibnamefont
  {van Elst}}, \bibinfo {author} {\bibfnamefont {J.}~\bibnamefont {Murugan}}, \
  and\ \bibinfo {author} {\bibfnamefont {J.-P.}\ \bibnamefont {Uzan}},\ }\href
  {\doibase 10.1088/0264-9381/28/22/225007} {\bibfield  {journal} {\bibinfo
  {journal} {Classical and Quantum Gravity}\ }\textbf {\bibinfo {volume}
  {28}},\ \bibinfo {pages} {225007} (\bibinfo {year} {2011})}\BibitemShut
  {NoStop}%
\bibitem [{\citenamefont {Josset}\ \emph {et~al.}(2017)\citenamefont {Josset},
  \citenamefont {Perez},\ and\ \citenamefont {Sudarsky}}]{Sudarsky}%
  \BibitemOpen
  \bibfield  {author} {\bibinfo {author} {\bibfnamefont {T.}~\bibnamefont
  {Josset}}, \bibinfo {author} {\bibfnamefont {A.}~\bibnamefont {Perez}}, \
  and\ \bibinfo {author} {\bibfnamefont {D.}~\bibnamefont {Sudarsky}},\ }\href
  {\doibase 10.1103/PhysRevLett.118.021102} {\bibfield  {journal} {\bibinfo
  {journal} {Phys. Rev. Lett.}\ }\textbf {\bibinfo {volume} {118}},\ \bibinfo
  {pages} {021102} (\bibinfo {year} {2017})}\BibitemShut {NoStop}%
\bibitem [{\citenamefont {Mitras}(2012)}]{Mitras}%
  \BibitemOpen
  \bibfield  {author} {\bibinfo {author} {\bibfnamefont {A.}~\bibnamefont
  {Mitras}},\ }\href {\doibase doi.org/10.1038/srep00923} {\bibfield  {journal}
  {\bibinfo  {journal} {Scientific Reports.}\ }\textbf {\bibinfo {volume} {2}}
  (\bibinfo {year} {2012}),\ doi.org/10.1038/srep00923}\BibitemShut {NoStop}%
\bibitem [{\citenamefont {Aghanim}\ \emph
  {et~al.}(2020{\natexlab{b}})\citenamefont {Aghanim} \emph
  {et~al.}}]{Planck:2019nip}%
  \BibitemOpen
  \bibfield  {author} {\bibinfo {author} {\bibfnamefont {N.}~\bibnamefont
  {Aghanim}} \emph {et~al.} (\bibinfo {collaboration} {Planck}),\ }\href
  {\doibase 10.1051/0004-6361/201936386} {\bibfield  {journal} {\bibinfo
  {journal} {Astron. Astrophys.}\ }\textbf {\bibinfo {volume} {641}},\ \bibinfo
  {pages} {A5} (\bibinfo {year} {2020}{\natexlab{b}})},\ \Eprint
  {http://arxiv.org/abs/1907.12875} {arXiv:1907.12875 [astro-ph.CO]}
  \BibitemShut {NoStop}%
\bibitem [{\citenamefont {Riess}\ \emph {et~al.}(2021)\citenamefont {Riess},
  \citenamefont {Casertano}, \citenamefont {Yuan}, \citenamefont {Bowers},
  \citenamefont {Macri}, \citenamefont {Zinn},\ and\ \citenamefont
  {Scolnic}}]{Riess:2020fzl}%
  \BibitemOpen
  \bibfield  {author} {\bibinfo {author} {\bibfnamefont {A.~G.}\ \bibnamefont
  {Riess}}, \bibinfo {author} {\bibfnamefont {S.}~\bibnamefont {Casertano}},
  \bibinfo {author} {\bibfnamefont {W.}~\bibnamefont {Yuan}}, \bibinfo {author}
  {\bibfnamefont {J.~B.}\ \bibnamefont {Bowers}}, \bibinfo {author}
  {\bibfnamefont {L.}~\bibnamefont {Macri}}, \bibinfo {author} {\bibfnamefont
  {J.~C.}\ \bibnamefont {Zinn}}, \ and\ \bibinfo {author} {\bibfnamefont
  {D.}~\bibnamefont {Scolnic}},\ }\href {\doibase 10.3847/2041-8213/abdbaf}
  {\bibfield  {journal} {\bibinfo  {journal} {Astrophys. J. Lett.}\ }\textbf
  {\bibinfo {volume} {908}},\ \bibinfo {pages} {L6} (\bibinfo {year} {2021})},\
  \Eprint {http://arxiv.org/abs/2012.08534} {arXiv:2012.08534 [astro-ph.CO]}
  \BibitemShut {NoStop}%
\bibitem [{\citenamefont {{Penzias}}\ and\ \citenamefont
  {{Wilson}}(1965)}]{PenziasAndWilson}%
  \BibitemOpen
  \bibfield  {author} {\bibinfo {author} {\bibfnamefont {A.~A.}\ \bibnamefont
  {{Penzias}}}\ and\ \bibinfo {author} {\bibfnamefont {R.~W.}\ \bibnamefont
  {{Wilson}}},\ }\href {\doibase 10.1086/148307} {\bibfield  {journal}
  {\bibinfo  {journal} {\apj}\ }\textbf {\bibinfo {volume} {142}},\ \bibinfo
  {pages} {419} (\bibinfo {year} {1965})}\BibitemShut {NoStop}%
\bibitem [{\citenamefont {De~Angelis}\ \emph {et~al.}(2013)\citenamefont
  {De~Angelis}, \citenamefont {Galanti},\ and\ \citenamefont
  {Roncadelli}}]{DeAngelis}%
  \BibitemOpen
  \bibfield  {author} {\bibinfo {author} {\bibfnamefont {A.}~\bibnamefont
  {De~Angelis}}, \bibinfo {author} {\bibfnamefont {G.}~\bibnamefont {Galanti}},
  \ and\ \bibinfo {author} {\bibfnamefont {M.}~\bibnamefont {Roncadelli}},\
  }\href {\doibase {10.1093/mnras/stt684}} {\bibfield  {journal} {\bibinfo
  {journal} {Monthly Notices of the Royal Astronomical Society}\ }\textbf
  {\bibinfo {volume} {432}},\ \bibinfo {pages} {3245} (\bibinfo {year}
  {2013})},\ \Eprint {http://arxiv.org/abs/1302.6460} {arXiv:1302.6460
  [astro-ph.HE]} \BibitemShut {NoStop}%
\bibitem [{\citenamefont {Gilmore}\ \emph {et~al.}(2012)\citenamefont
  {Gilmore}, \citenamefont {Somerville}, \citenamefont {Primack},\ and\
  \citenamefont {Dominguez}}]{Gilmore:2011ks}%
  \BibitemOpen
  \bibfield  {author} {\bibinfo {author} {\bibfnamefont {R.}~\bibnamefont
  {Gilmore}}, \bibinfo {author} {\bibfnamefont {R.}~\bibnamefont {Somerville}},
  \bibinfo {author} {\bibfnamefont {J.}~\bibnamefont {Primack}}, \ and\
  \bibinfo {author} {\bibfnamefont {A.}~\bibnamefont {Dominguez}},\ }\href
  {\doibase 10.1111/j.1365-2966.2012.20841.x} {\bibfield  {journal} {\bibinfo
  {journal} {Mon. Not. Roy. Astron. Soc.}\ }\textbf {\bibinfo {volume} {422}},\
  \bibinfo {pages} {3189} (\bibinfo {year} {2012})},\ \Eprint
  {http://arxiv.org/abs/1104.0671} {arXiv:1104.0671 [astro-ph.CO]} \BibitemShut
  {NoStop}%
\bibitem [{\citenamefont {Franceschini}\ \emph {et~al.}(2008)\citenamefont
  {Franceschini}, \citenamefont {Rodighiero},\ and\ \citenamefont
  {Vaccari}}]{Franceschini:2008tp}%
  \BibitemOpen
  \bibfield  {author} {\bibinfo {author} {\bibfnamefont {A.}~\bibnamefont
  {Franceschini}}, \bibinfo {author} {\bibfnamefont {G.}~\bibnamefont
  {Rodighiero}}, \ and\ \bibinfo {author} {\bibfnamefont {M.}~\bibnamefont
  {Vaccari}},\ }\href {\doibase 10.1051/0004-6361:200809691} {\bibfield
  {journal} {\bibinfo  {journal} {Astron. Astrophys.}\ }\textbf {\bibinfo
  {volume} {487}},\ \bibinfo {pages} {837} (\bibinfo {year} {2008})},\ \Eprint
  {http://arxiv.org/abs/0805.1841} {arXiv:0805.1841 [astro-ph]} \BibitemShut
  {NoStop}%
\bibitem [{\citenamefont {Dominguez}\ \emph {et~al.}(2011)\citenamefont
  {Dominguez} \emph {et~al.}}]{Dominguez:2010bv}%
  \BibitemOpen
  \bibfield  {author} {\bibinfo {author} {\bibfnamefont {A.}~\bibnamefont
  {Dominguez}} \emph {et~al.},\ }\href {\doibase
  10.1111/j.1365-2966.2010.17631.x} {\bibfield  {journal} {\bibinfo  {journal}
  {Mon. Not. Roy. Astron. Soc.}\ }\textbf {\bibinfo {volume} {410}},\ \bibinfo
  {pages} {2556} (\bibinfo {year} {2011})},\ \Eprint
  {http://arxiv.org/abs/1007.1459} {arXiv:1007.1459 [astro-ph.CO]} \BibitemShut
  {NoStop}%
\bibitem [{\citenamefont {Breit}\ and\ \citenamefont
  {Wheeler}(1934)}]{Breit:1934zz}%
  \BibitemOpen
  \bibfield  {author} {\bibinfo {author} {\bibfnamefont {G.}~\bibnamefont
  {Breit}}\ and\ \bibinfo {author} {\bibfnamefont {J.~A.}\ \bibnamefont
  {Wheeler}},\ }\href {\doibase 10.1103/PhysRev.46.1087} {\bibfield  {journal}
  {\bibinfo  {journal} {Phys. Rev.}\ }\textbf {\bibinfo {volume} {46}},\
  \bibinfo {pages} {1087} (\bibinfo {year} {1934})}\BibitemShut {NoStop}%
\bibitem [{\citenamefont {Weidenspointner}\ \emph {et~al.}(2000)\citenamefont
  {Weidenspointner}, \citenamefont {Varendorff}, \citenamefont {Kappadath},
  \citenamefont {Bennett}, \citenamefont {Bloemen}, \citenamefont {Diehl},
  \citenamefont {Hermsen}, \citenamefont {Lichti}, \citenamefont {Ryan},\ and\
  \citenamefont {Schönfelder}}]{COMPTEL01}%
  \BibitemOpen
  \bibfield  {author} {\bibinfo {author} {\bibfnamefont {G.}~\bibnamefont
  {Weidenspointner}}, \bibinfo {author} {\bibfnamefont {M.}~\bibnamefont
  {Varendorff}}, \bibinfo {author} {\bibfnamefont {S.~C.}\ \bibnamefont
  {Kappadath}}, \bibinfo {author} {\bibfnamefont {K.}~\bibnamefont {Bennett}},
  \bibinfo {author} {\bibfnamefont {H.}~\bibnamefont {Bloemen}}, \bibinfo
  {author} {\bibfnamefont {R.}~\bibnamefont {Diehl}}, \bibinfo {author}
  {\bibfnamefont {W.}~\bibnamefont {Hermsen}}, \bibinfo {author} {\bibfnamefont
  {G.~G.}\ \bibnamefont {Lichti}}, \bibinfo {author} {\bibfnamefont
  {J.}~\bibnamefont {Ryan}}, \ and\ \bibinfo {author} {\bibfnamefont
  {V.}~\bibnamefont {Schönfelder}},\ }\href {\doibase 10.1063/1.1307028}
  {\bibfield  {journal} {\bibinfo  {journal} {AIP Conference Proceedings}\
  }\textbf {\bibinfo {volume} {510}},\ \bibinfo {pages} {467} (\bibinfo {year}
  {2000})}\BibitemShut {NoStop}%
\bibitem [{\citenamefont {Strong}\ \emph {et~al.}(2004)\citenamefont {Strong},
  \citenamefont {Moskalenko},\ and\ \citenamefont {Reimer}}]{EGRET01}%
  \BibitemOpen
  \bibfield  {author} {\bibinfo {author} {\bibfnamefont {A.~W.}\ \bibnamefont
  {Strong}}, \bibinfo {author} {\bibfnamefont {I.~V.}\ \bibnamefont
  {Moskalenko}}, \ and\ \bibinfo {author} {\bibfnamefont {O.}~\bibnamefont
  {Reimer}},\ }\href {\doibase 10.1086/423196} {\bibfield  {journal} {\bibinfo
  {journal} {Astrophys. J.}\ }\textbf {\bibinfo {volume} {613}},\ \bibinfo
  {pages} {956} (\bibinfo {year} {2004})},\ \Eprint
  {http://arxiv.org/abs/astro-ph/0405441} {arXiv:astro-ph/0405441} \BibitemShut
  {NoStop}%
\bibitem [{\citenamefont {Ajello}\ \emph {et~al.}(2008)\citenamefont {Ajello},
  \citenamefont {Greiner}, \citenamefont {Sato}, \citenamefont {Willis},
  \citenamefont {Kanbach}, \citenamefont {Strong}, \citenamefont {Diehl},
  \citenamefont {Hasinger}, \citenamefont {Gehrels}, \citenamefont
  {Markwardt},\ and\ \citenamefont {Tueller}}]{Ajello_2008}%
  \BibitemOpen
  \bibfield  {author} {\bibinfo {author} {\bibfnamefont {M.}~\bibnamefont
  {Ajello}}, \bibinfo {author} {\bibfnamefont {J.}~\bibnamefont {Greiner}},
  \bibinfo {author} {\bibfnamefont {G.}~\bibnamefont {Sato}}, \bibinfo {author}
  {\bibfnamefont {D.~R.}\ \bibnamefont {Willis}}, \bibinfo {author}
  {\bibfnamefont {G.}~\bibnamefont {Kanbach}}, \bibinfo {author} {\bibfnamefont
  {A.~W.}\ \bibnamefont {Strong}}, \bibinfo {author} {\bibfnamefont
  {R.}~\bibnamefont {Diehl}}, \bibinfo {author} {\bibfnamefont
  {G.}~\bibnamefont {Hasinger}}, \bibinfo {author} {\bibfnamefont
  {N.}~\bibnamefont {Gehrels}}, \bibinfo {author} {\bibfnamefont {C.~B.}\
  \bibnamefont {Markwardt}}, \ and\ \bibinfo {author} {\bibfnamefont
  {J.}~\bibnamefont {Tueller}},\ }\href {\doibase 10.1086/592595} {\bibfield
  {journal} {\bibinfo  {journal} {The Astrophysical Journal}\ }\textbf
  {\bibinfo {volume} {689}},\ \bibinfo {pages} {666} (\bibinfo {year}
  {2008})}\BibitemShut {NoStop}%
\bibitem [{\citenamefont {Oberlack}(2010)}]{Oberlack:2010}%
  \BibitemOpen
  \bibfield  {author} {\bibinfo {author} {\bibfnamefont {U.}~\bibnamefont
  {Oberlack}},\ }\href {\doibase 10.1103/Physics.3.21} {\bibfield  {journal}
  {\bibinfo  {journal} {Physics}\ }\textbf {\bibinfo {volume} {3}} (\bibinfo
  {year} {2010}),\ 10.1103/Physics.3.21}\BibitemShut {NoStop}%
\bibitem [{\citenamefont {{Akaike}}(1974)}]{Akaike:1974}%
  \BibitemOpen
  \bibfield  {author} {\bibinfo {author} {\bibfnamefont {H.}~\bibnamefont
  {{Akaike}}},\ }\href@noop {} {\bibfield  {journal} {\bibinfo  {journal} {IEEE
  Transactions on Automatic Control}\ }\textbf {\bibinfo {volume} {19}},\
  \bibinfo {pages} {716} (\bibinfo {year} {1974})}\BibitemShut {NoStop}%
\bibitem [{\citenamefont {Sugiura}(1978)}]{Sugiura:1978}%
  \BibitemOpen
  \bibfield  {author} {\bibinfo {author} {\bibfnamefont {N.}~\bibnamefont
  {Sugiura}},\ }\href {\doibase 10.1080/03610927808827599} {\bibfield
  {journal} {\bibinfo  {journal} {Communications in Statistics - Theory and
  Methods}\ }\textbf {\bibinfo {volume} {7}},\ \bibinfo {pages} {13} (\bibinfo
  {year} {1978})}\BibitemShut {NoStop}%
\bibitem [{\citenamefont {Aramaki}\ \emph {et~al.}(2020)\citenamefont
  {Aramaki}, \citenamefont {Adrian}, \citenamefont {Karagiorgi},\ and\
  \citenamefont {Odaka}}]{ARAMAKI2020107}%
  \BibitemOpen
  \bibfield  {author} {\bibinfo {author} {\bibfnamefont {T.}~\bibnamefont
  {Aramaki}}, \bibinfo {author} {\bibfnamefont {P.~O.~H.}\ \bibnamefont
  {Adrian}}, \bibinfo {author} {\bibfnamefont {G.}~\bibnamefont {Karagiorgi}},
  \ and\ \bibinfo {author} {\bibfnamefont {H.}~\bibnamefont {Odaka}},\ }\href
  {\doibase https://doi.org/10.1016/j.astropartphys.2019.07.002} {\bibfield
  {journal} {\bibinfo  {journal} {Astroparticle Physics}\ }\textbf {\bibinfo
  {volume} {114}},\ \bibinfo {pages} {107} (\bibinfo {year}
  {2020})}\BibitemShut {NoStop}%
\bibitem [{\citenamefont {Shutt}\ \emph {et~al.}(2023)\citenamefont {Shutt},
  \citenamefont {Akerib}, \citenamefont {Breur}, \citenamefont {Buuck},
  \citenamefont {Dragone}, \citenamefont {Digel}, \citenamefont {Haller},
  \citenamefont {Hitchcock}, \citenamefont {Linehan}, \citenamefont {Luitz},
  \citenamefont {Madejski}, \citenamefont {Monzani}, \citenamefont {Petrillo},
  \citenamefont {Pivovaroff}, \citenamefont {Tanaka}, \citenamefont
  {Tompkins},\ and\ \citenamefont {A}}]{GammaTPC}%
  \BibitemOpen
  \bibfield  {author} {\bibinfo {author} {\bibfnamefont {T.}~\bibnamefont
  {Shutt}}, \bibinfo {author} {\bibfnamefont {D.}~\bibnamefont {Akerib}},
  \bibinfo {author} {\bibfnamefont {S.}~\bibnamefont {Breur}}, \bibinfo
  {author} {\bibfnamefont {M.}~\bibnamefont {Buuck}}, \bibinfo {author}
  {\bibfnamefont {A.}~\bibnamefont {Dragone}}, \bibinfo {author} {\bibfnamefont
  {S.}~\bibnamefont {Digel}}, \bibinfo {author} {\bibfnamefont
  {G.}~\bibnamefont {Haller}}, \bibinfo {author} {\bibfnamefont
  {O.}~\bibnamefont {Hitchcock}}, \bibinfo {author} {\bibfnamefont
  {R.}~\bibnamefont {Linehan}}, \bibinfo {author} {\bibfnamefont
  {S.}~\bibnamefont {Luitz}}, \bibinfo {author} {\bibfnamefont
  {G.}~\bibnamefont {Madejski}}, \bibinfo {author} {\bibfnamefont
  {M.}~\bibnamefont {Monzani}}, \bibinfo {author} {\bibfnamefont
  {G.}~\bibnamefont {Petrillo}}, \bibinfo {author} {\bibfnamefont
  {M.}~\bibnamefont {Pivovaroff}}, \bibinfo {author} {\bibfnamefont
  {H.}~\bibnamefont {Tanaka}}, \bibinfo {author} {\bibfnamefont
  {L.}~\bibnamefont {Tompkins}}, \ and\ \bibinfo {author} {\bibfnamefont
  {Y.-T.~T.}\ \bibnamefont {A}},\ }\href@noop {} {\  (\bibinfo {year} {last
  access November, 2023})},\ \url
  {https://www.snowmass21.org/docs/files/summaries/CF/SNOWMASS21-CF7_CF1-NF7_NF10-IF8_IF0_Shutt-224.pdf} \BibitemShut {NoStop}%
\bibitem [{\citenamefont {Garc\'ia-Aspeitia}\ \emph {et~al.}(2019)\citenamefont
  {Garc\'ia-Aspeitia}, \citenamefont {Mart\'inez-Robles}, \citenamefont
  {Hern\'andez-Almada}, \citenamefont {Maga\~na},\ and\ \citenamefont
  {Motta}}]{Garcia-Aspeitia:2019yni}%
  \BibitemOpen
  \bibfield  {author} {\bibinfo {author} {\bibfnamefont {M.~A.}\ \bibnamefont
  {Garc\'ia-Aspeitia}}, \bibinfo {author} {\bibfnamefont {C.}~\bibnamefont
  {Mart\'inez-Robles}}, \bibinfo {author} {\bibfnamefont {A.}~\bibnamefont
  {Hern\'andez-Almada}}, \bibinfo {author} {\bibfnamefont {J.}~\bibnamefont
  {Maga\~na}}, \ and\ \bibinfo {author} {\bibfnamefont {V.}~\bibnamefont
  {Motta}},\ }\href {\doibase 10.1103/PhysRevD.99.123525} {\bibfield  {journal}
  {\bibinfo  {journal} {Phys. Rev.}\ }\textbf {\bibinfo {volume} {D99}},\
  \bibinfo {pages} {123525} (\bibinfo {year} {2019})},\ \Eprint
  {http://arxiv.org/abs/1903.06344} {arXiv:1903.06344 [gr-qc]} \BibitemShut
  {NoStop}%
\bibitem [{\citenamefont {Garc\'\i{}a-Aspeitia}\ \emph
  {et~al.}(2021)\citenamefont {Garc\'\i{}a-Aspeitia}, \citenamefont
  {Hern\'andez-Almada}, \citenamefont {Maga\~na},\ and\ \citenamefont
  {Motta}}]{Garcia-Aspeitia:2019yod}%
  \BibitemOpen
  \bibfield  {author} {\bibinfo {author} {\bibfnamefont {M.~A.}\ \bibnamefont
  {Garc\'\i{}a-Aspeitia}}, \bibinfo {author} {\bibfnamefont {A.}~\bibnamefont
  {Hern\'andez-Almada}}, \bibinfo {author} {\bibfnamefont {J.}~\bibnamefont
  {Maga\~na}}, \ and\ \bibinfo {author} {\bibfnamefont {V.}~\bibnamefont
  {Motta}},\ }\href {\doibase 10.1016/j.dark.2021.100840} {\bibfield  {journal}
  {\bibinfo  {journal} {Phys. Dark Univ.}\ }\textbf {\bibinfo {volume} {32}},\
  \bibinfo {pages} {100840} (\bibinfo {year} {2021})},\ \Eprint
  {http://arxiv.org/abs/1912.07500} {arXiv:1912.07500 [astro-ph.CO]}
  \BibitemShut {NoStop}%
\bibitem [{\citenamefont {Zhang}\ \emph {et~al.}(2017)\citenamefont {Zhang},
  \citenamefont {Li},\ and\ \citenamefont {Xia}}]{Zhang:2016}%
  \BibitemOpen
  \bibfield  {author} {\bibinfo {author} {\bibfnamefont {M.-J.}\ \bibnamefont
  {Zhang}}, \bibinfo {author} {\bibfnamefont {H.}~\bibnamefont {Li}}, \ and\
  \bibinfo {author} {\bibfnamefont {J.-Q.}\ \bibnamefont {Xia}},\ }\href
  {\doibase 10.1140/epjc/s10052-017-5005-4} {\bibfield  {journal} {\bibinfo
  {journal} {Eur. Phys. J.}\ }\textbf {\bibinfo {volume} {C77}},\ \bibinfo
  {pages} {434} (\bibinfo {year} {2017})},\ \Eprint
  {http://arxiv.org/abs/1601.01758} {arXiv:1601.01758 [astro-ph.CO]}
  \BibitemShut {NoStop}%
\bibitem [{\citenamefont {{Zel'dovich}}\ and\ \citenamefont
  {{Novikov}}(1966)}]{Zeldovich1966}%
  \BibitemOpen
  \bibfield  {author} {\bibinfo {author} {\bibfnamefont {Y.~B.}\ \bibnamefont
  {{Zel'dovich}}}\ and\ \bibinfo {author} {\bibfnamefont {I.~D.}\ \bibnamefont
  {{Novikov}}},\ }\href@noop {} {\bibfield  {journal} {\bibinfo  {journal}
  {Astronomicheskii Zhurnal}\ }\textbf {\bibinfo {volume} {43}},\ \bibinfo
  {pages} {758} (\bibinfo {year} {1966})}\BibitemShut {NoStop}%
\bibitem [{\citenamefont {{Hawking}}(1971)}]{Hawking1971}%
  \BibitemOpen
  \bibfield  {author} {\bibinfo {author} {\bibfnamefont {S.}~\bibnamefont
  {{Hawking}}},\ }\href {\doibase 10.1093/mnras/152.1.75} {\bibfield  {journal}
  {\bibinfo  {journal} {Monthly Notices of the Royal Astronomical Society}\
  }\textbf {\bibinfo {volume} {152}},\ \bibinfo {pages} {75} (\bibinfo {year}
  {1971})}\BibitemShut {NoStop}%
\bibitem [{\citenamefont {Carr}\ and\ \citenamefont
  {Hawking}(1974)}]{Carr_Hawking:1974}%
  \BibitemOpen
  \bibfield  {author} {\bibinfo {author} {\bibfnamefont {B.~J.}\ \bibnamefont
  {Carr}}\ and\ \bibinfo {author} {\bibfnamefont {S.~W.}\ \bibnamefont
  {Hawking}},\ }\href {\doibase 10.1093/mnras/168.2.399} {\bibfield  {journal}
  {\bibinfo  {journal} {Monthly Notices of the Royal Astronomical Society}\
  }\textbf {\bibinfo {volume} {168}},\ \bibinfo {pages} {399} (\bibinfo {year}
  {1974})},\ \Eprint
  {http://arxiv.org/abs/https://academic.oup.com/mnras/article-pdf/168/2/399/8079885/mnras168-0399.pdf}
  {https://academic.oup.com/mnras/article-pdf/168/2/399/8079885/mnras168-0399.pdf}
  \BibitemShut {NoStop}%
\bibitem [{\citenamefont {Gonzalez-Morales}\ \emph {et~al.}(2017)\citenamefont
  {Gonzalez-Morales}, \citenamefont {Profumo},\ and\ \citenamefont
  {Reynoso-C\'ordova}}]{Gonzalez-Morales:2017jkx}%
  \BibitemOpen
  \bibfield  {author} {\bibinfo {author} {\bibfnamefont {A.~X.}\ \bibnamefont
  {Gonzalez-Morales}}, \bibinfo {author} {\bibfnamefont {S.}~\bibnamefont
  {Profumo}}, \ and\ \bibinfo {author} {\bibfnamefont {J.}~\bibnamefont
  {Reynoso-C\'ordova}},\ }\href {\doibase 10.1103/PhysRevD.96.063520}
  {\bibfield  {journal} {\bibinfo  {journal} {Phys. Rev. D}\ }\textbf {\bibinfo
  {volume} {96}},\ \bibinfo {pages} {063520} (\bibinfo {year} {2017})},\
  \Eprint {http://arxiv.org/abs/1705.00777} {arXiv:1705.00777 [astro-ph.CO]}
  \BibitemShut {NoStop}%
\bibitem [{\citenamefont {Boudaud}\ \emph {et~al.}(2019)\citenamefont
  {Boudaud}, \citenamefont {Lacroix}, \citenamefont {Stref},\ and\
  \citenamefont {Lavalle}}]{Boudaud:2018oya}%
  \BibitemOpen
  \bibfield  {author} {\bibinfo {author} {\bibfnamefont {M.}~\bibnamefont
  {Boudaud}}, \bibinfo {author} {\bibfnamefont {T.}~\bibnamefont {Lacroix}},
  \bibinfo {author} {\bibfnamefont {M.}~\bibnamefont {Stref}}, \ and\ \bibinfo
  {author} {\bibfnamefont {J.}~\bibnamefont {Lavalle}},\ }\href {\doibase
  10.1103/PhysRevD.99.061302} {\bibfield  {journal} {\bibinfo  {journal} {Phys.
  Rev. D}\ }\textbf {\bibinfo {volume} {99}},\ \bibinfo {pages} {061302}
  (\bibinfo {year} {2019})},\ \Eprint {http://arxiv.org/abs/1810.01680}
  {arXiv:1810.01680 [astro-ph.HE]} \BibitemShut {NoStop}%
\bibitem [{\citenamefont {Diósi}(1987)}]{Diosi1}%
  \BibitemOpen
  \bibfield  {author} {\bibinfo {author} {\bibfnamefont {L.}~\bibnamefont
  {Diósi}},\ }\href {\doibase https://doi.org/10.1016/0375-9601(87)90681-5}
  {\bibfield  {journal} {\bibinfo  {journal} {Physics Letters A}\ }\textbf
  {\bibinfo {volume} {120}},\ \bibinfo {pages} {377} (\bibinfo {year}
  {1987})}\BibitemShut {NoStop}%
\bibitem [{\citenamefont {Di\'osi}(1989)}]{Diosi2}%
  \BibitemOpen
  \bibfield  {author} {\bibinfo {author} {\bibfnamefont {L.}~\bibnamefont
  {Di\'osi}},\ }\href {\doibase 10.1103/PhysRevA.40.1165} {\bibfield  {journal}
  {\bibinfo  {journal} {Phys. Rev. A}\ }\textbf {\bibinfo {volume} {40}},\
  \bibinfo {pages} {1165} (\bibinfo {year} {1989})}\BibitemShut {NoStop}%
\bibitem [{\citenamefont {Penrose}(1996)}]{Penrose}%
  \BibitemOpen
  \bibfield  {author} {\bibinfo {author} {\bibfnamefont {R.}~\bibnamefont
  {Penrose}},\ }\href {\doibase 10.1007/BF02105068} {\bibfield  {journal}
  {\bibinfo  {journal} {General Relativity and Gravitation}\ }\textbf {\bibinfo
  {volume} {28}},\ \bibinfo {pages} {581} (\bibinfo {year} {1996})}\BibitemShut
  {NoStop}%
\bibitem [{\citenamefont {Peccei}(2008)}]{Axions}%
  \BibitemOpen
  \bibfield  {author} {\bibinfo {author} {\bibfnamefont {R.~D.}\ \bibnamefont
  {Peccei}},\ }\href {\doibase 10.1007/978-3-540-73518-2_1} {\bibfield
  {journal} {\bibinfo  {journal} {Lect. Notes Phys.}\ }\textbf {\bibinfo
  {volume} {741}},\ \bibinfo {pages} {3} (\bibinfo {year} {2008})},\ \Eprint
  {http://arxiv.org/abs/hep-ph/0607268} {arXiv:hep-ph/0607268} \BibitemShut
  {NoStop}%
\bibitem [{\citenamefont {Essig}\ \emph {et~al.}(2013)\citenamefont {Essig}
  \emph {et~al.}}]{DarkPhotons}%
  \BibitemOpen
  \bibfield  {author} {\bibinfo {author} {\bibfnamefont {R.}~\bibnamefont
  {Essig}} \emph {et~al.},\ }in\ \href@noop {} {\emph {\bibinfo {booktitle}
  {Snowmass 2013: Snowmass on the}}} {\emph {\bibinfo {booktitle}
  {Mississippi}}} \ (\bibinfo {year} {2013})\
  \Eprint {http://arxiv.org/abs/1311.0029} {arXiv:1311.0029 [hep-ph]}
  \BibitemShut {NoStop}%
\end{thebibliography}

%

\end{document}